\documentclass[12pt]{article}
\usepackage{epsf}
\usepackage{amsmath, amssymb}
\usepackage{graphicx}
\usepackage{comment,color}
\usepackage{cite}

\def\a{\alpha}
\def\b{\beta}

\def\d{\delta}
\def\e{\epsilon}

\def\m{\mu}
\def\n{\nu}

\def\q{\partial}

\def\s{\sigma}

\def\L{\Lambda}

\begin{document}

\newcommand{\bi}{\bfseries\itshape}

\renewcommand{\thefootnote}{\fnsymbol{footnote}}
\setcounter{footnote}{0}
\begin{titlepage}

\def\thefootnote{\fnsymbol{footnote}}
\begin{flushright}
KEK-TH-1688
\end{flushright}
\begin{center}

\vskip .75in

{\Large \bf 
 Composite Higgsino 
}

\vskip .75in

{\large
Masaki Asano$^{(a)}$ and Ryosuke Sato$^{(b)}$ 
}

\vskip 0.25in

{\em $^{(a)}$
Physikalisches Institut and Bethe Center for Theoretical Physics, \\ 
Universit\"{a}t Bonn, Nussallee 12, D-53115 Bonn, Germany
}

\vskip 0.1in

{\em $^{(b)}$ Theory Center, KEK, 1-1 Oho, Tsukuba, Ibaraki 305-0801, Japan
}

\end{center}
\vskip .5in

\begin{abstract}

Several supersymmetric models in which there is a (partially) composite Higgs boson arising from (coupled to) a strong sector have been proposed. Such strong dynamics would help to cause the electroweak symmetry breaking naturally. In this paper, we focus on the compositeness of the Higgsinos in such models. We show that such a Higgsino compositeness can induce the characteristic decay branching fraction of the neutralinos. In such scenarios, the decay branching fraction of the second lightest neutral Higgsino into the lightest neutral Higgsino with photon can be large due to a dipole interaction. We also discuss the Higgsino dark matter feature. The annihilation cross section into $\gamma Z$ can be large. 

\end{abstract}

\end{titlepage}

\renewcommand{\thepage}{\arabic{page}}
\setcounter{page}{1}
\renewcommand{\thefootnote}{\#\arabic{footnote}}
\setcounter{footnote}{0}

\section{Introduction}

Large Hadron collider (LHC) experiments are searching for new physics beyond the standard model (SM) related with the electroweak symmetry breaking and low-energy supersymmetry (SUSY) is one of the leading candidate. For example, in the minimal supersymmetric standard model (MSSM), the electroweak symmetry breaking scale is regarded as the combination of supersymmetric mass ($\mu$ term) and soft SUSY breaking parameters. 

In order to achieve the electroweak symmetry breaking naturally, each contribution would not be much larger than the observed electroweak scale because if these are large, big cancellation between these contributions is required. 
In particular, because the main contribution to the electroweak symmetry breaking is arising from the $\mu$ term and up-type Higgs soft mass $m_{H_u}$ in MSSM, a small value ($\mathcal{O}(100)$ GeV) of these parameters 
is favored by such a naturalness discussion. 
Moreover, stop mass and the $A$ term, $A_t$, are also constrained from the naturalness discussion, because the $m_{H_u}$ receives the stop loop correction 
$\delta m_{H_u}^2 \sim - 3 y_t^2/(8 \pi^2) (m_{Q_3}^2 + m_{U_3}^2 + |A_t|^2) \ln (M_{\rm mess}/m_{\tilde{t}})$ 
at weak scale
\footnote{
For details of such a ``natural supersymmetry" spectrum, e.g., see Refs. \cite{Kitano:2006gv, Asano:2010ut, Papucci:2011wy}. 
}.

However, the stop masses and $A_t$ are also related with the lightest neutral Higgs boson mass $m_h$ via the loop diagram; the heavy stop lifts the Higgs mass up. Thus, in order to avoid the fine-tuning with $m_h = 125$ GeV \cite{Aad:2012tfa}\cite{Chatrchyan:2012ufa}, models which receives additional contributions to the Higgs mass and the low messenger scale $M_{\rm mess}$would be favored \cite{NMSSMmirage}.

One of the way to construct such a natural supersymmetric spectrum is introducing a strong dynamics in low-energy SUSY scenario (e.g., partially composite Higgs models \cite{Fukushima:2010pm,Craig:2011ev,Csaki:2011xn,Azatov:2011ht,Azatov:2011ps,Gherghetta:2011na,Heckman:2011bb,
 Csaki:2012fh,Evans:2012uf,Kitano:2012wv,Barbieri:2013hxa}). 
\footnote{ 
For earlier proposals, see also \cite{Witten:1981DBS,Dine:1981ST,Dimopoulos:1981au}\cite{Samuel:1990dq,Dine:1990jd,Harnik:2003rs}.
}
If Higgs fields are in (or coupled to) a strong sector, the additional contribution to the Higgs potential will be arising from the strong sector dynamically. In composite Higgs scenario, the potential is generated at the dynamical scale, thus, $M_{\rm mess}$ can be taken to be small. Moreover, it can also be considered that the Higgs soft mass is suppressed by entering the strong sector into the superconformal window in partially composite Higgs scenarios. If it is suppressed enough, $M_{\rm mess}$ can be small also in partially composite Higgs scenarios, effectively.

In particular, after the discovery of the 125 GeV Higgs boson which property is similar to the SM (i.e. elementary) Higgs boson, the scenario with the partially composite Higgs boson in low-energy SUSY became more attractive. Such a property of the Higgs boson is promised by supersymmetry and the naturalness of the scale would implies an existence of the strong dynamics, additionally.

In this paper, we focus on the compositeness of the Higgsino in such SUSY scenarios. The (partially) composite Higgsino has magnetic moment couplings arising from the strong dynamics. Then, the branching ratio of the Higgsino decays into the lightest Higgsino can drastically change. Actually, we show that the decay of the second lightest neutralino into the lightest neutralino will dominantly accommodate with photon in the case where the composite Higgsino composes the lightest neutralino dominantly. 

Finally, we also discuss the other possibilities to appear the Higgsino compositeness. In particular, we show that the magnetic moment can also play an important role to search and constrain the (partially) composite Higgsino dark matter.

\section{A (partially) composite Higgsino}
In models with (partially) composite Higgs, 
the effect of strong dynamics can be written as higher dimensional
operators, and the following term gives anomalous magnetic couplings 
\footnote{ 
For the anomalous magnetic moment in a supersymmetric case, see Refs. \cite{Ferrara:1974wb, Ferrara:1992nm}.  
},
\begin{eqnarray}
 \int d^2 \theta \int d^2 \bar{\theta} H_d W_\a e^V (D^\a e^{-V} H_u) \mathcal{Z}^\dagger, 
   \label{eq:Com_SS0}
\end{eqnarray}
where $H_u$($H_d$) and $W_\a$ are chiral superfield for the up(down)-type Higgs doublet and field-strength superfield for the $SU(2)$ or $U(1)$ gauge symmetry in the SM, respectively. 
The $D_\a$ is the derivative operator in superspace and $V$ denotes the SM vector superfields. The $\mathcal{Z}^\dagger$ depends on the SUSY soft breaking mass, $\mathcal{Z}^\dagger = \bar{\theta}^2 m_{\rm SUSY}$.

This includes the following magnetic moment interaction and also kinetic mixing terms;
\begin{eqnarray}
\mathcal{L} \supset c_{\rm dipole} \bar{\tilde{H}}_d \sigma_{\mu\nu} \tilde{H}_u F^{\mu\nu} 
+ c_{\rm mixing} H_d \tilde{H}_u \s^\m \q_\m \tilde{W}^\dagger, 
   \label{eq:Com_SS1}
\end{eqnarray}
where the $\tilde{H}_u$($\tilde{H}_d$) are the fermion component of $H_u$($H_d$) and the $F^{\mu\nu}$ is the field strength for the SM $SU(2)$ or $U(1)$ gauge symmetry. The $c_{\rm dipole}$ and $c_{\rm mixing}$ in Eq. (\ref{eq:Com_SS1}) are constants and its values depend on models. 
As we will see later, these terms, the dipole term especially, can change the Higgsino phenomenology drastically.

In this section, at first, we describe the general form of these terms using naive dimensional analysis. Then, we introduce several scenarios motivated by current experimental results and the benchmark values of the coefficient of the dipole interaction $c_{\rm dipole}$ in Eq. (\ref{eq:Com_SS1}) briefly.

\subsection{General description for Composite Higgsino case}

At first, we consider the dipole term of the composite Higgsino in general, considering the Lagrangian by naive dimensional analysis \cite{Manohar:1983md,Georgi:1986kr,Georgi:1992dw,Luty:1997fk,Cohen:1997rt}, 
\begin{eqnarray}
\mathcal{L}_{\rm NDA} 
&=& \frac{\Lambda^4}{g_\rho^2} \int d^2 \theta \int d^2 \bar{\theta}  
    \Lambda^{-2} \mathcal{K} 
    \left( \frac{g_\rho \epsilon_{u,d} H_{u,d}}{\Lambda}, 
           \frac{D}{\Lambda^{1/2}}, 
           \frac{g W}{\Lambda^{3/2}}, ... \right)
\nonumber \\ && 
   + \frac{\Lambda^4}{g_\rho^2} 
     \left[ \int d^2 \theta \Lambda^{-1} \mathcal{W} 
            \left( \frac{g_\rho \epsilon_{u,d} H_{u,d}}{\Lambda}, 
                   \frac{D}{\Lambda^{1/2}}, 
                   \frac{g W}{\Lambda^{3/2}}, ... 
           \right) + {\rm h.c.} \right], 
   \label{eq:SUSY_NDA}
\end{eqnarray}
where $g_\rho \sim 4 \pi$ and $g$ is the SM gauge coupling. The $\Lambda$ is the dynamical scale; the heavy sector are integrated out and we obtain the effective theory at the scale. Here, we parametrize the compositeness of $H_u$($H_d$) using $\epsilon_u$($\epsilon_d$) and $\epsilon_{u,d} \sim 1$ is corresponding to the fully composite $H_{u,d}$ case.

From Eq. (\ref{eq:SUSY_NDA}), we obtained the following term, 
\begin{eqnarray}
\mathcal{L}_{\rm NDA} 
\supset \mathcal{L}_{\rm NDA}^{\rm dipole} 
&=& \int d^2 \theta \int d^2 \bar{\theta} ~ 
        g \frac{ \epsilon_u \epsilon_d }{\Lambda^2} 
        H_d W_\a e^V (D^\a e^{-V} H_u) \mathcal{Z}^\dagger  + {\rm h.c.} 
\nonumber \\
&\sim& 
        g \left( \frac{ \epsilon_u \epsilon_d }{\Lambda} \right) 
          \left( \frac{m_{\rm SUSY}}{\Lambda} \right)
        \left[ H_d W_\a e^V (D^\a e^{-V} H_u) \right] |_{\theta^2}  + {\rm h.c.}, 
   \label{eq:comp_general}
\end{eqnarray}
Then, the dipole term can be obtained as 
\begin{eqnarray}
\mathcal{L}_{\rm NDA}^{\rm dipole} 
&\supset&
 g c_{\rm soft} \frac{ \epsilon_u \epsilon_d }{2\Lambda}
 \bar{\tilde{H}}_d \sigma_{\mu\nu} \tilde{H}_u F^{\mu\nu}, 
   \label{eq:def_c}
\end{eqnarray}
where $c_{\rm soft} = m_{\rm SUSY} / \Lambda$. 
Thus, the $c_{\rm dipole}$ in Eq. (\ref{eq:Com_SS1}) can be written by $c_{\rm dipole} \sim g c_{\rm soft} (\epsilon_u \epsilon_d/\Lambda)$.

Furthermore, it also includes the following term, 
\begin{eqnarray}
\mathcal{L}_{\rm NDA}^{\rm dipole} 
&\supset&
 g c_{\rm soft} \frac{\sqrt{2}i \epsilon_u \epsilon_d }{\Lambda} 
        H_d \tilde{H}_u \s^\m \q_\m \tilde{W}^\dagger, 
   \label{eq:comp_kineticmix}
\end{eqnarray}
and it contributes to the kinetic mixing of neutralinos and charginos. We describe such kinetic mixing terms and the correction of the neutralino mixing matrix in Appendix. We will discuss the effect which comes from this correction later. For example, it can change the cross section related with dark matter direct detection experiments.

\subsection{Partially composite Higgs model}

Here, we discuss the low-energy SUSY scenario with composite Higgs and the benchmark values of $c_{\rm dipole}$ in proposed models briefly. In particular, we focus on the scenario which has partially composite Higgs bosons. In the scenario, because there is also elementary Higgs, the SM fermion masses can be arising from the Yukawa couplings without additional flavor problems as usual in MSSM. 
\footnote{
But, of course, the results of our study can be useful for all scenario which include a composite Higgsino or partially composite Higgsino taking the $c_{\rm dipole}$ and $c_{\rm mixing}$ value of the scenario. 
}

We consider such a partially composite Higgsino case supposing the following superpotential, 
\begin{eqnarray}
\mathcal{W} = \lambda_d H_d \mathcal{O}_u +  \lambda_u H_u \mathcal{O}_d, 
   \label{eq:c_PartComp}
\end{eqnarray}
where elementary superfields $H_{u,d}$ interact with the strong sector via the coupling $\lambda_{u(d)}$ has the mass dimension $2-d$ and the $d$ is the dimension of a composite operator $\mathcal{O}_{u(d)}$. These can be correspond with $\epsilon_{u,d}$ in Eq. (\ref{eq:SUSY_NDA}); 
\begin{eqnarray}
\epsilon_{u,d} = \frac{\lambda_{u,d}}{\Lambda^{2-d}},  
   \label{eq:epsilon_lambda}
\end{eqnarray}
where $\epsilon_{u,d}$ are dimensionless parameters and $\Lambda$ is regarded as the dynamical scale induced the strong sector. Since the Higgs boson is partially composite, the strong sector also contributes to generate the Higgs potential, then, the $125$ GeV mass of the lightest neutral Higgs boson can be explain without very large SUSY breaking soft masses.

Considering the strong superconformal sector as the strong sector and $d < 2$, the couplings in Eq. (\ref{eq:c_PartComp}) are relevant. Even for the large $\lambda_{u,d}$ which is required to explain the $125$ GeV Higgs mass, Landau poles in the UV can be avoided.

The dynamically generated $\mu$ term can be a desirable value and a solution to the $\mu$ problem in low-energy SUSY. In this scenario, it is required that the $\lambda_{u,d}^{1/(2-d)}$ is around the TeV scale in order to generate the $125$ GeV Higgs mass, then, it is possible to cause a coincidence problem. However, the coincidence problem can be solved by, at least, considering an extension of the Giudice-Masiero mechanism \cite{Azatov:2011ht,Azatov:2011ps,Kitano:2012wv}.

To obtain a natural supersymmetric spectrum, it would be also required some mechanism due to null results of current LHC SUSY searches, e.g., to obtain a hierarchy between the colored SUSY particle masses and the Higgs soft mass. Also in order to obtain that, the superconformal sector may be helpful. In the scenario, the Higgs soft mass can be affected by the superconformal feature, then, the suppressed Higgs soft mass could be obtained at the dynamical scale 
\cite{Nelson:2000sn,Nelson:2001mq,Kobayashi:2001kz,Luty:2001jh,Luty:2001zv,Dine:2004dv,Ibe:2005pj,Ibe:2005qv,Schmaltz:2006qs,Cohen:2006qc}.

In the partially composite Higgs boson scenarios, there are two possibility to break the electroweak symmetry breaking; the electroweak symmetry breaking vacuum expectation value (VEV) is composed also by strong sector (case 1) or only by elementary Higgs VEV (case 2). And there are several possibilities to achieve non-zero vacuum expectation values of Higgs fiels; for example, the minimization is dominated by balancing between the $H^2$ term and the smaller power-law term or between the $H^2$ term and the larger power-law term. However, in order to explain the smallness of the electroweak symmetry breaking scale, $\Lambda$ should not be much larger than $1$ TeV and, actually, the values of $\epsilon_u \epsilon_d / \Lambda$ are not much different from $\sim 1/(\mathcal{O}(10) {\rm TeV})$ in proposed scenarios. For more details about models, see papers of each scenarios, e.g., Refs \cite{Fukushima:2010pm,Craig:2011ev,Csaki:2011xn,Azatov:2011ht,Azatov:2011ps,Gherghetta:2011na,Heckman:2011bb,
 Csaki:2012fh,Evans:2012uf,Kitano:2012wv}. On the other hand, the $c_{\rm soft}$ value which parametrize also the SUSY breaking in the strong sector fully depends on scenario.

\section{Signal in collider}

In this section, we demonstrate the effect of the dipole term to the branching fraction of the neutralino decay. At first, we show the decay branching ratio of the second lightest neutralino $\tilde{\chi}^0_2$ in the gaugino decoupling limit, $M_1$, $M_2 \gg \mu$, $m_Z$ (where $M_2$($M_1$) is the $SU(2)$($U(1)$) gaugino mass), for simplicity. Here, we consider the following magnetic moment coupling, 
\begin{eqnarray}
\mathcal{L} 
&=& \frac{e}{\Lambda_{\rm dipole}} 
\bar{\tilde{\chi}}^0_1 \sigma_{\mu\nu} \tilde{\chi}^0_2 F^{\mu\nu}, 
   \label{eq:MMcoupling}
\end{eqnarray}
as the contribution from UV theory. The $\Lambda_{\rm dipole}$ can be reinterpreted as 
$1/\Lambda_{\rm dipole} \sim c_{\rm soft} \epsilon_u \epsilon_d/ \Lambda$ 
using $\Lambda$ and $\epsilon_{u,d}$, in Eq. (\ref{eq:SUSY_NDA}).

In MSSM, the $\tilde{\chi}_2^0$ can decay into $\tilde{\chi}_1^0 \gamma$ via loop diagrams. The decay width of such a two body decay is written by 
\begin{eqnarray}
\Gamma_{\tilde{\chi}^0_2 \to \tilde{\chi}^0_1 \gamma}^{\rm MSSM}
&\sim& [C(m_W^2/\m^2 )]^2 \frac{\alpha_{\rm em}^3}{4 \pi^2 \sin^4\theta_W} 
           \frac{(m_{\tilde{\chi}^0_2} - m_{\tilde{\chi}^0_1})^3}{\mu^2}
\nonumber \\
&\sim& [C(m_W^2/\m^2 )]^2 \frac{\alpha_{\rm em}^3}{4 \pi^2 \sin^4\theta_W} \frac{m_Z^6}{\mu^2}
           \left(
                  \frac{\sin^2\theta_W}{M_1} + \frac{\cos^2\theta_W}{M_2}
          \right)^3, 
   \label{eq:width2body_MSSM}
\end{eqnarray}
where $C(r)$  can be obtained by calculating the integrals explicitly.
For small $r \ll 1$, $C(r) = (\log r)/2 + 1 - 3\pi r^{1/2} / 4 + {\cal O}(r\log r)$. See Ref.\cite{Haber:1988px} for an explicit formula of $C(r)$.

On the other hand, the decay width of the three body decay via virtual $Z$ boson exchange is written by 
\begin{eqnarray}
\sum_f \Gamma_{\tilde{\chi}^0_2 \to \tilde{\chi}^0_1 f \bar{f}} 
&\sim& \frac{\alpha_{\rm em}^2}{30 \pi \cos^4\theta_W}  m_Z^6 
       \left( \frac{\sin^2\theta_W}{M_1} + \frac{\cos^2\theta_W}{M_2} \right)^5
       \left( \frac{40}{3} - \frac{10}{\sin^2\theta_W} + \frac{21}{4 \sin^4\theta_W} \right). 
   \label{eq:width2body}
\end{eqnarray}
Here, $f$ is summed over all of the quarks and leptons expect for top quark. 
Then, in MSSM, the branching ratio of the two-body decay to the three-body decay is obtained as 
\begin{eqnarray}
\frac{ \Gamma_{\tilde{\chi}^0_2 \to \tilde{\chi}^0_1 \gamma}^{\rm MSSM} } 
     { \sum_f \Gamma_{\tilde{\chi}^0_2 \to \tilde{\chi}^0_1 f \bar{f}} } 
&\sim& [C(m_W^2/\m^2 )]^2
       \frac{15 \alpha_{\rm em}}{2 \pi \sin^4\theta_W} \frac{1}{\mu^2}
       \left( \frac{M_1 M_2}{M_1 + M_2 \tan^2\theta_W} \right)^2
\nonumber \\ 
&& \qquad \qquad \qquad \qquad \qquad \quad \times 
       \left( \frac{40}{3} - \frac{10}{\sin^2\theta_W} + \frac{21}{4 \sin^4\theta_W} \right)^{-1}, 
   \label{eq:ratio_2body3body_MSSM}
\end{eqnarray}
thus, the branching fraction of $\tilde{\chi}_2^0 \to \tilde{\chi}_1^0 \gamma$ become large at $M_{\rm gaugino} \gg |\mu|$ limit in MSSM \cite{Haber:1988px}.

Next, we estimate the contribution of the dipole term in Eq. (\ref{eq:MMcoupling}) induced from strong dynamics. The two body decay width from the magnetic dipole moment coupling is obtained as 
\begin{eqnarray}
\Gamma_{\tilde{\chi}^0_2 \to \tilde{\chi}^0_1 \gamma}^{\rm dipole} 
&=& \frac{1}{8\pi} \left( \frac{2 e}{\Lambda_{\rm dipole}} \right)^2 
    \frac{(m_{\tilde{\chi}^0_2}^2 - m_{\tilde{\chi}^0_1}^2)^3}{m_{\tilde{\chi}^0_2}^3}
\nonumber \\ 
&\sim& \frac{16 \alpha_{\rm em}}{\Lambda_{\rm dipole}^2} m_Z^6 
       \left( \frac{\sin^2\theta_W}{M_1} + \frac{\cos^2\theta_W}{M_2} \right)^3. 
\end{eqnarray}
where we calculate only the dipole coupling contribution neglecting the MSSM one-loop diagram. And the ratio is written by 
\begin{eqnarray}
\frac{ \Gamma_{\tilde{\chi}^0_2 \to \tilde{\chi}^0_1 \gamma}^{\rm dipole}  } 
     { \sum_f \Gamma_{\tilde{\chi}^0_2 \to \tilde{\chi}^0_1 f \bar{f}} } 
&\sim& \frac{480 \pi}{\alpha_{\rm em}}  \frac{1}{\Lambda_{\rm dipole}^2} 
       \left( \frac{M_1 M_2}{M_1 + M_2 \tan^2\theta_W} \right)^2
       \left( \frac{40}{3} - \frac{10}{\sin^2\theta_W} + \frac{21}{4 \sin^4\theta_W} \right)^{-1}.
\end{eqnarray}
This shows that the contribution from the dipole term would be comparable with the MSSM loop contribution in a parameter region with $\mu \sim 300$ GeV and $\Lambda_{\rm dipole} \sim \Lambda / (c_{\rm soft} \epsilon_u \epsilon_d) \sim 250$ TeV. Furthermore, in this limit, the branching ratio of the two-body decay to three-body decay is the same order when $M_1 \sim M_2 \sim 1$ TeV and $\Lambda_{\rm dipole} \sim \mathcal{O}(100)$ TeV.

\begin{figure}[htbp]
  \begin{center}
    \includegraphics[width=13.5cm]{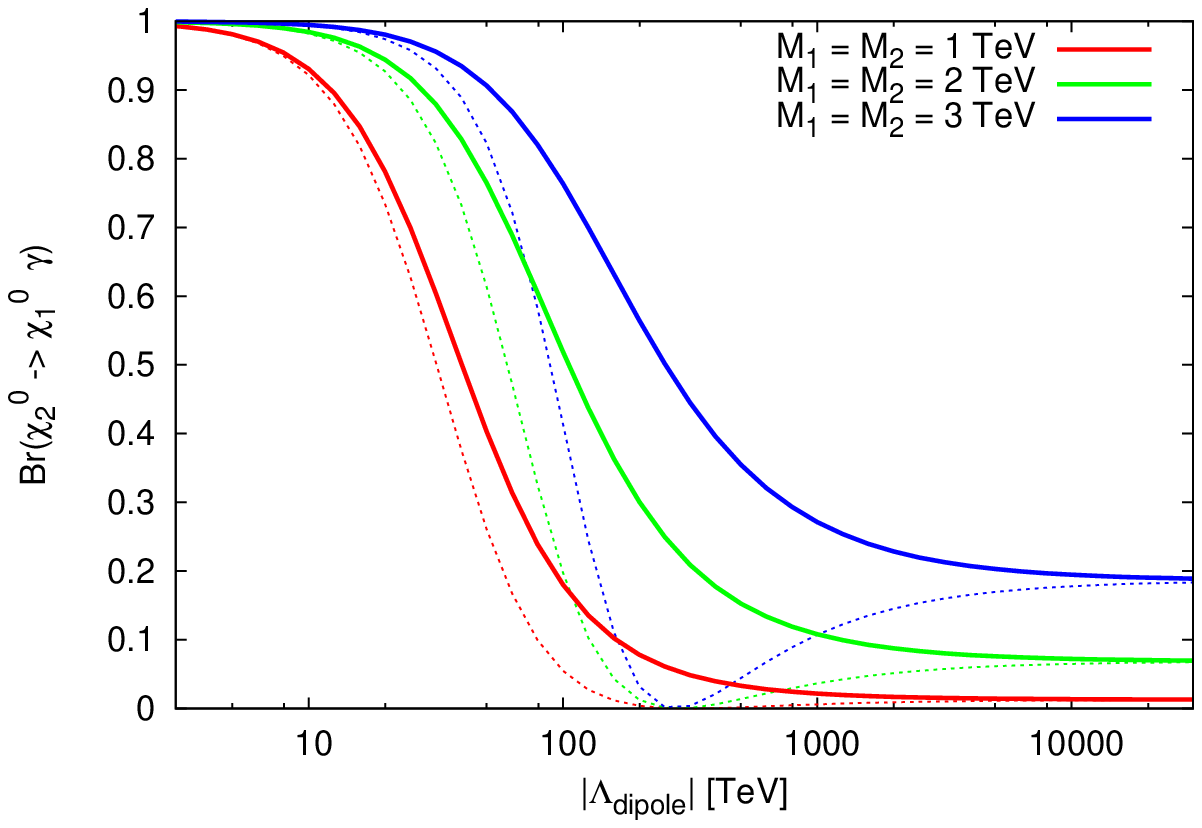}
    \includegraphics[width=13.5cm]{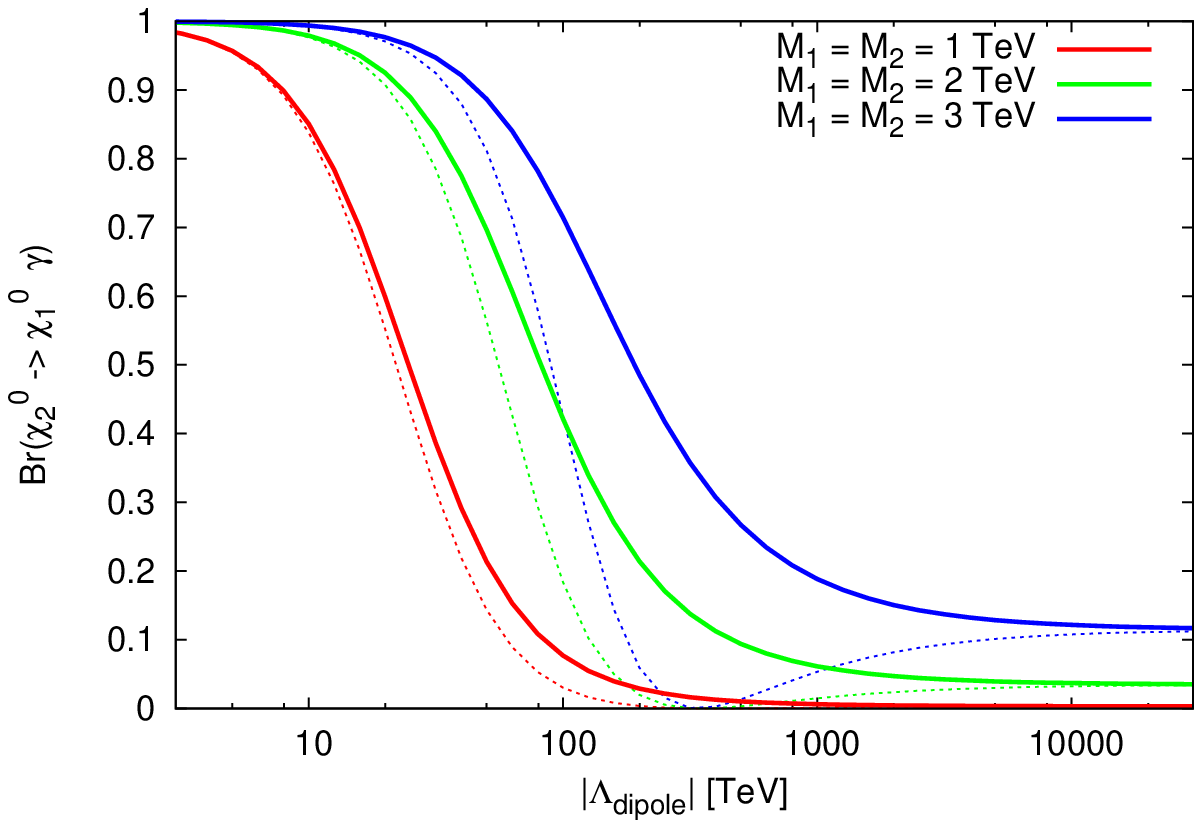}
    \caption{\small The branching fraction of the $\tilde{\chi}_2^0 \to \tilde{\chi}_1^0 \gamma$ decay which depends on $|\Lambda_{\rm dipole}| \sim |\Lambda / (c_{\rm soft} \epsilon_u \epsilon_d)|$. 
 The $\mu$ term is $\mu = 200$ GeV ($500$ GeV) in the upper (lower) figure.
Each solid line shows the case in which the gaugino mass for $SU(2)$ and $U(1)$ are the same
and the value is $1$ TeV (red), $2$ TeV (green) and $3$ TeV (blue) with positive $\Lambda_{\rm dipole}$, and each dashed line shows the case with negative $\Lambda_{\rm dipole}$. 
Here, we take $\tan\b = 5$ and $\m>0$.
              }
   \label{fig:Fig_ratio}
  \end{center}
\end{figure}

Finally we show the parameter dependence of the branching fraction of $\tilde{\chi}_2^0$ decay for more details. Here, we use ISAJET783 \cite{Paige:2003mg} with a modification to include the additional dipole contribution. The Fig.\ref{fig:Fig_ratio} shows the branching ratio of the two-body decay in the second-lightest neutralino decay. In the calculation, we take the same soft mass for Bino and Wino, decoupled squark and slepton masses, and assume that the neutralino mass matrix is the same as the MSSM, for simplicity. Although the components, actually, can be slightly different from the MSSM, the quantitative feature of the neutralino decay is not drastically changed. 
\begin{figure}[htbp]
  \begin{center}
    \includegraphics[width=13.5cm]{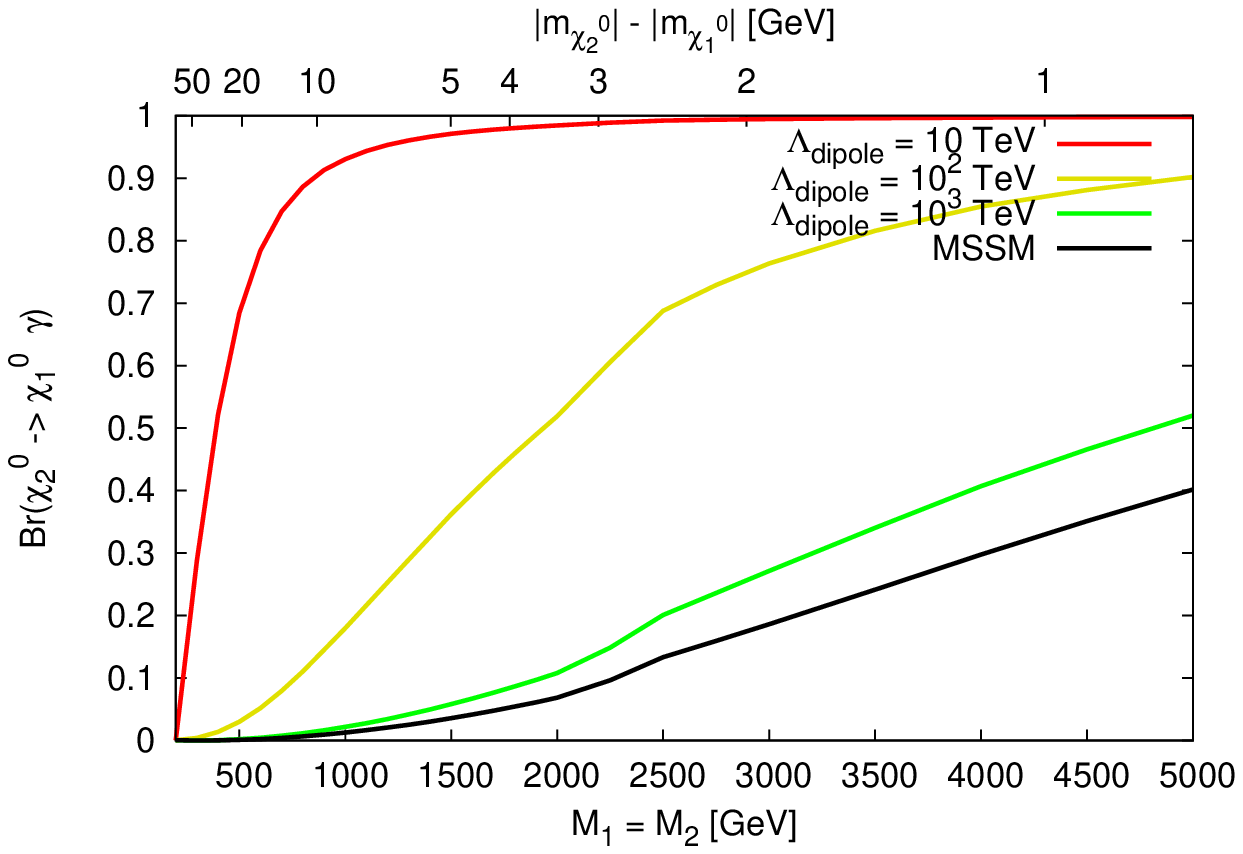}
    \includegraphics[width=13.5cm]{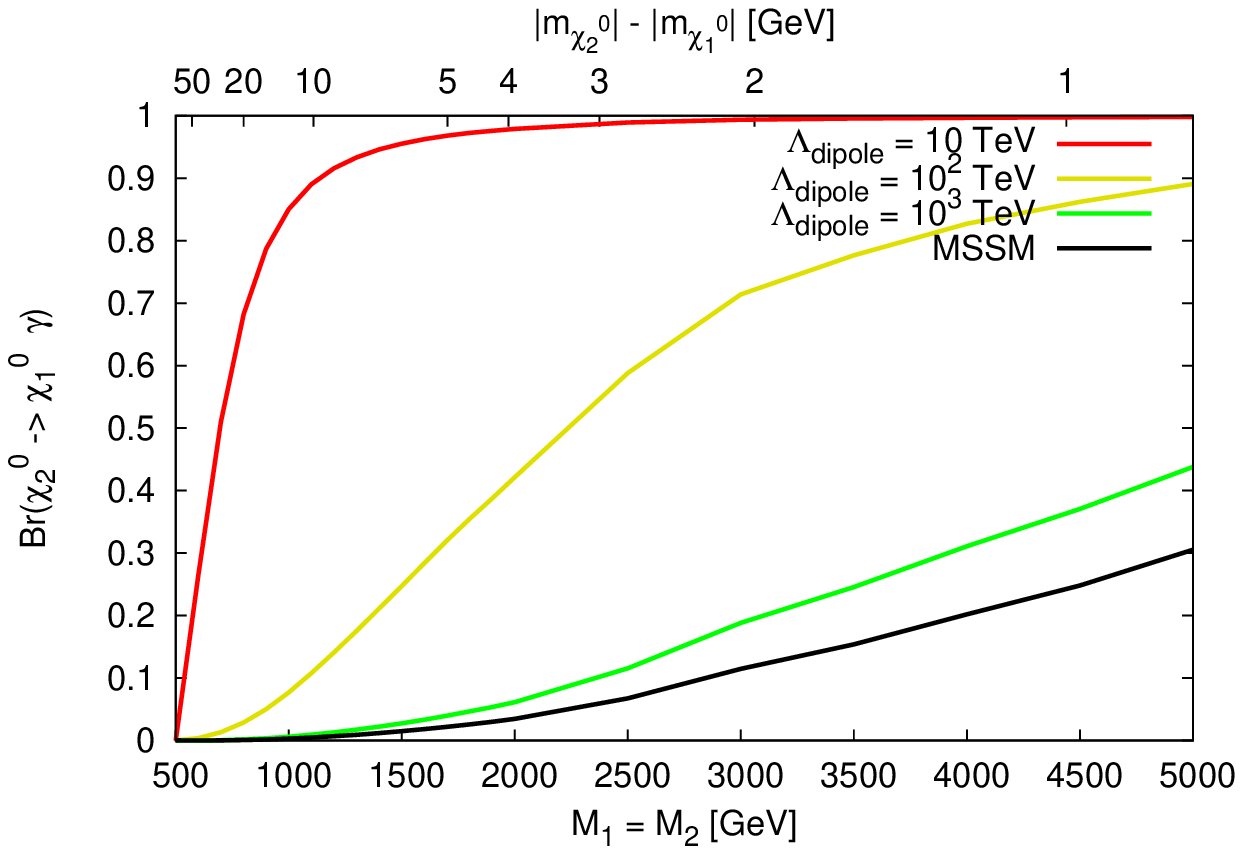}
    \caption{\small The branching fraction of the $\tilde{\chi}_2^0 \to \tilde{\chi}_1^0 \gamma$ decay which 
depends on $|\Lambda_{\rm dipole}| \sim |\Lambda / (c_{\rm soft} \epsilon_u \epsilon_d)|$. 
The $\mu$ term is $\mu = 200$ GeV ($500$ GeV) in the upper (lower) figure.  We take the same values for $SU(2)$ and $U(1)$ gaugino masses
and each line shows the case in which the $\Lambda_{\rm dipole} = 10$ TeV (red), $100$ TeV (yellow), $1000$ TeV (green) and $\infty$ TeV (black). 
The limit of $\L_{\rm dipole} = \infty$ TeV corresponds to the MSSM.
Here, we take $\tan\b = 5$, $\m>0$ and $\L_{\rm dipole} > 0$.
            }
   \label{fig:Fig_ratio_2}
  \end{center}
\end{figure}

As shown in Fig.\ref{fig:Fig_ratio}, the larger $\Lambda_{\rm dipole}$, the smaller the fraction of the $\tilde{\chi}_2^0 \to \tilde{\chi}_1^0 \gamma$ branch. But, at $|\Lambda_{\rm dipole}| \sim |\Lambda / (c_{\rm soft} \epsilon_u \epsilon_d)| \sim \mathcal{O}(100)$ TeV, the contribution from the higher dimensional operator is comparable with the contribution from MSSM one-loop diagrams. And, in a case where the dipole term has a negative sign, $\Lambda_{\rm dipole} < 0$, the contribution cancel the MSSM loop contribution at such a region. Then, in very large $|\Lambda_{\rm dipole}|$ region, the branching ratio is a constant in the Fig.\ref{fig:Fig_ratio}. The Fig.\ref{fig:Fig_ratio} shows that, for example, in $\mu = 200$ and $M_1 = M_2 =1$ TeV case, the branching fraction of the $\tilde{\chi}_2^0 \to \tilde{\chi}_1^0 \gamma$ is greater than $90 \%$ 
if $|\Lambda_{\rm dipole}| \sim |\Lambda / (c_{\rm soft} \epsilon_u \epsilon_d)| = 10$ TeV. 

We also show the gaugino mass dependence in Fig.\ref{fig:Fig_ratio_2}. It can be seen that the branching fraction of $\tilde{\chi}_2^0 \to \tilde{\chi}_1^0 \gamma$ can be sizable even in light gaugino mass region if a large contribution from the higher dimensional term exists in the composite Higgs scenario.

\section{Signal in space}
So far, we investigated the branching ratio of the Higgsino which can be changed by the dipole term arising from the strong sector. In this section, we discuss other possibilities to appear the compositeness of Higgsinos. 

\begin{figure}[t]
  \begin{center}
    \includegraphics[width=6cm]{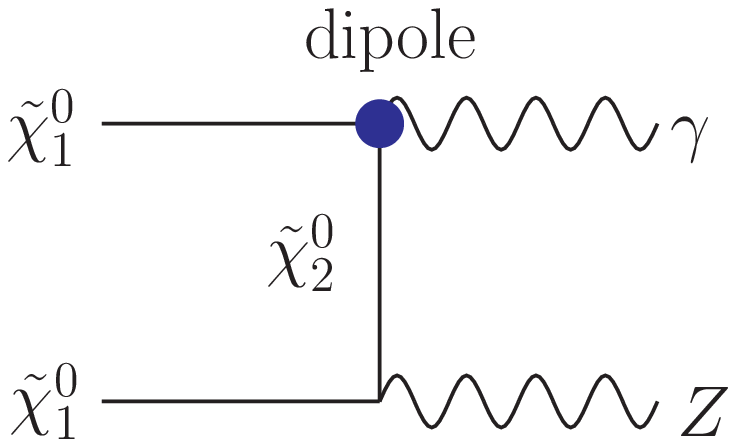}
    \caption{\small Diagrams of Higgsino annihilation cross section included dipole interaction into $\gamma Z$. 
            }
   \label{fig:Fig_anni_diagram}
  \end{center}
\end{figure}

The dipole term can also change the feature of the Higgsino dark matter. The Higgsino-like dark matter can annihilate into $\gamma \gamma$ and $\gamma Z$ via loop diagrams and the annihilation cross section is $(\sigma{\rm v})_{\gamma \gamma (\gamma Z)} \sim 1.0 (2.2) \times 10^{-28}$ cm$^3$s$^{-1}$ at $\mu \sim 140$ GeV pure Higgsino region in MSSM \cite{Bergstrom:1997fh}\cite{Ullio:1997ke}\cite{Boudjema:2005hb}
\footnote{We have calculated these values using the micromegas \cite{Belanger:2013oya}. }. 
On the other hand, It would be possible that these cross section become large in the (partially) composite Higgsino case; the cross section which induced by the diagram shown in Fig.\ref{fig:Fig_anni_diagram} can be $(\sigma{\rm v})_{\gamma Z}^{\rm dipole} \sim 10^{-27}$ cm$^3$s$^{-1}$ at $|\Lambda_{\rm dipole}| \sim |\Lambda / (c_{\rm soft} \epsilon_u \epsilon_d)| \sim 10$ TeV. 

 Thus, the cross section is around the current limit by Fermi \cite{Fermi-LAT:2013uma} and a comparable order of a line-like spectral feature which was reported by Refs.\cite{Bringmann:2012vr,Weniger:2012tx} using the data of the Fermi Gamma-Ray Space Telescope \cite{Atwood:2009ez}. (See also Ref.\cite{Bringmann:2012ez} and references therein for details of this signal. The current Fermi-LAT reports the local (global) significance of $3.3$ ($1.5$) $\sigma$ at $133$ GeV \cite{Fermi-LAT:2013uma}.) 

In addition to the line-gamma constraints, we should also take into account the constraints from continuum gamma-ray and other cosmic-ray observations \cite{Buckley:2012ws,Buchmuller:2012rc,Cohen:2012me,Cholis:2012fb,Asano:2012zv} for the Higgsino dark matter scenario because of the large annihilate cross section to weak gauge bosons, $(\sigma{\rm v})_{WW (ZZ)} \sim 2.1 (1.8) \times 10^{-25}$ cm$^3$s$^{-1}$ at $\mu \sim 140$ GeV pure Higgsino region in MSSM \cite{Olive:1990qm}. But, there are also uncertainties and because the ratio of the $WW$($ZZ$) cross section to $\gamma Z$ cross section is not determined, unlike the wino and Higgsino case in MSSM \cite{Buchmuller:2012rc}, the possibility for the composite Higgsino as the origin of the tentative gamma-line like signal cannot be excluded by a dark matter distribution independent way.

There are also another possibilities to appear the compositeness of Higgsinos. As we mentioned in previous section, the off-diagonal component in neutralino and chargino mass matrices can receive the corrections due to $v_1^2 + v_2^2 \ne (246 {\rm ~GeV})^2$ in case 1 introduced in Section 2. And the kinetic mixing term also change the neutralino and chargino mass matrices (for details, see Appendix). These correction can change the cross section of the Higgsino dark matter in direct detection experiments; If gaugino masses are large, $m_Z/M_{1(2)} \sim m_Z/\Lambda_{\rm dipole} \sim c_{\rm soft} \epsilon_u \epsilon_d m_Z/\Lambda$, the effect cannot be neglected.

These cosmological and astrophysical aspects of the (partially) composite Higgsino are very interesting and important, but, the detail analysis is beyond the scope of this paper. 
\footnote{
For other features of this dark matter, e.g., see also Refs. \cite{Cheung:2009qk,Berg:2009mq,Bernal:2009hd,Bernal:2009jc}. 
}
And, regardless of whether the Higgsino is dark matter or not, the decay branching ratio of the second-lightest neutral Higgsino can be changed as we shown in this paper. We will study the details for the dark matter phenomenology in other place \cite{preparation}.

\section{Summary}

In this paper, we investigate the possibility to appear the Higgsino compositeness in the neutralino decay. In the partially composite Higgs scenarios which are motivated by explaining the electroweak symmetry breaking naturally, the dipole term of the Higgsinos would be arising from the strong sector and it can change the branching ratio of the neutralino. As shown in section 3, the $\tilde{\chi}^0_2 \to \tilde{\chi}^0_1 \gamma$ branch can be dominant at a plausible parameter space.

Furthermore, we also discuss other possibilities to appear the compositeness of Higgsinos in the previous section. We show that the possible corrections for the dark matter feature not only from the dipole terms but also from the kinetic mixing terms. In particular, the dipole terms can change the annihilation cross section of Higgsino to $\gamma Z$ drastically. Although there is $\mathcal{O}(1)$ uncertainties in our estimation and also the model dependent factor $c_{\rm soft}$, it can contribute to constraint for the dark matter scenario in the partially composite Higgs models. And, in some case in which $c_{\rm soft} \sim 1$, such large annihilation cross section into $\gamma Z$ could help to explain the tentative gamma-line like signal in the Fermi-LAT. 

Even from a view point of dynamical electroweak symmetry breaking scenarios with composite Higgs boson, the existence of the elementary Higgs are also attractive because Yukawa couplings can be written without additional flavor problems. Then, to protect the mass squared term of the elementary Higgs, supersymmetry is a viable symmetry.

In several scenario of the partially composite Higgs scenarios, the Higgs physics can deviate from the SM. But, the current experimental results are, unfortunately, consistent with the SM Higgs boson. However, the Higgsinos are also partially composite in such models. 
\footnote{
And there are also TeV resonances in this scenario. These could also be discovered at LHC \cite{Kitano:2012wv}. 
}
Thus, the indication of the compositeness due to a (semi)-perturbative coupling with a strong sector could also be measured at the Higgsino phenomenology.

\section*{Acknowledgments}
We would like to thank R. Kitano and E. Dudas for discussions. 
M.A. acknowledges support from the German Research Foundation (DFG) through
grant BR 3954/1-1 and DFG TRR33 ``The Dark Universe". 
R.S. is grateful to DESY theory group for their hospitality during the course of this work.
The work of R.S. is supported in part by JSPS Research Fellowships for Young Scientists.

\appendix
\section*{Appendix}
In this Appendix, we describe the effects of the following operators in detail:
\begin{eqnarray}
{\cal O}_{B,1} = \int d^4 \theta \mathcal{Z}^\dagger H_d W^\a_Y e^V (D_\a e^{-V} H_u), 
\qquad 
{\cal O}_{B,2} = \int d^4 \theta \mathcal{Z}^\dagger H_u W^\a_Y e^V (D_\a e^{-V} H_d),
\\
{\cal O}_{W,1} = \int d^4 \theta \mathcal{Z}^\dagger H_d W^\a e^V (D_\a e^{-V} H_u), 
\qquad 
{\cal O}_{W,2} = \int d^4 \theta \mathcal{Z}^\dagger H_u W^\a e^V (D_\a e^{-V} H_d),
 \label{eq:operators}
\end{eqnarray}
where $W$($W_Y$) are the field strength superfields for $SU(2)_L$($U(1)_Y$) gauge symmetry and $V$ denotes the SM vector superfields, respectively. The $D_\alpha$ is the derivative operator in superspace. Superfields can be expanded by its component fields as the following,
\begin{eqnarray}
H_d &=& H_d + \sqrt{2} \theta {\tilde H}_d + \theta^2 F_{H_d},\\
H_u &=& H_u + \sqrt{2} \theta {\tilde H}_u + \theta^2 F_{H_u},\\
W_{Y\a} &=& {\tilde B}_\a + \theta_\a D_Y + \frac{1}{4} (\s^{\m\n} \theta )_\a B_{\m\n} + i \theta^2 (\s^\m \q_\m {\tilde B}^\dagger ),\\
W_\a &=& {\tilde W}_\a + \theta_\a D + \frac{1}{4} (\s^{\m\n} \theta )_\a W_{\m\n} + i \theta^2 (\s^\m \q_\m {\tilde W}^\dagger ). 
\end{eqnarray}
By using component fields, the SUSY breaking contribution in the operators can be written by, 
\begin{eqnarray}
{\cal O}_{B,1} &\supset& 
m_{\rm soft} \left[
\sqrt{2} F_{H_d} {\tilde B} {\tilde H}_u - \sqrt{2} F_{H_u} {\tilde B} {\tilde H}_d - D_Y {\tilde H}_u {\tilde H}_d \right . \nonumber\\
&& \left .
+ 2 H_d D_Y F_{H_u} + \sqrt{2} i H_d {\tilde H}_u \s^\m \q_\m {\tilde B}^\dagger - \frac{1}{4} {\tilde H}_u \s^{\m\n} {\tilde H}_d B_{\m\n}
\right], \label{eq:op1}\\
{\cal O}_{B,2} &\supset& 
m_{\rm soft} \left[
\sqrt{2} F_{H_u} {\tilde B} {\tilde H}_d - \sqrt{2} F_{H_d} {\tilde B} {\tilde H}_u - D_Y {\tilde H}_u {\tilde H}_d \right . \nonumber\\
&&  \left .
+ 2 H_u D_Y F_{H_d} + \sqrt{2} i H_u {\tilde H}_d \s^\m \q_\m {\tilde B}^\dagger - \frac{1}{4} {\tilde H}_d \s^{\m\n} {\tilde H}_u B_{\m\n}
\right]. \label{eq:op2}
\end{eqnarray}
The ${\cal O}_{W,1}$ and ${\cal O}_{W,2}$ can also be expanded in the same manner. From equations of motion, we get $F_{H_u} = \m v_d$, $F_{H_d} = \m v_u$,
$D_Y = -g'(v_u^2 -v_d^2)/4$ and $D_3 = g(v_u^2 -v_d^2)/4$. In the above operators, the first three terms contribute to the neutralino mass matrix, the fifth term gives the kinetic mixing between Higgsino and gauginos, and the last term gives the dipole operator for Higgsino
\footnote{
The supersymmetric part of the operators also includes the contribution to the mixing, although these are suppressed $\mathcal{O}(v^2/\Lambda^2)$. 
}. 

To demonstrate these contribution, we consider the following effective operators: 
\begin{eqnarray}
{\cal L}_{\rm eff.} &=& 
\frac{g'\e_u \e_d}{\L}C_{g',u} \left[ {\cal O}_{B,1} \right]_{\theta^2} 
+\frac{g'\e_u \e_d}{\L}C_{g',d} \left[ {\cal O}_{B,2} \right]_{\theta^2} 
\nonumber \\ && 
+\frac{g\e_u \e_d}{\L}C_{g,u} \left[ {\cal O}_{W,1} \right]_{\theta^2} 
-\frac{g\e_u \e_d}{\L}C_{g,d} \left[ {\cal O}_{W,2} \right]_{\theta^2} 
 + h.c., 
\end{eqnarray}
where $C_{g'(g),u(d)}$ is a coefficient which depends on the SUSY breaking as $C_{\rm soft}$ in Eq.(\ref{eq:def_c}). From now on, we discuss the effects of the above operators on mixing of the neutralinos at the order of $g'(g) \epsilon^2 C_{g'(g)}/\L$. We define canonically normalized neutralino fields $(\tilde B', \tilde W', {\tilde H}_d', {\tilde H}_u')$ as,
\begin{eqnarray}
\left(\begin{array}{c}
\tilde B\\
\tilde W^0\\
{\tilde H}_d \\
{\tilde H}_u
\end{array}\right)
=
\left( 1 + \d{\cal N}_{\rm kin.} \right)
\left(\begin{array}{c}
{\tilde B}'\\
{\tilde W}'\\
{\tilde H}_d ' \\
{\tilde H}_u '
\end{array}\right),
\end{eqnarray}
where $\d{\cal N}_{\rm kin.}$ is given as,
\begin{eqnarray}
\d {\cal N}_{\rm kin.}
&\simeq&
\frac{2\e_u \e_d m_Z}{\L}
\left(
\begin{array}{cccc}
0                 & 0                & C_{g',d} s_\b s_W & -C_{g',u} c_\b s_W \\
0                 & 0                & -C_{g,d}  s_\b c_W & C_{g,u} c_\b c_W\\
C_{g',d} s_\b s_W & -C_{g,d} s_\b c_W & 0                 & 0  \\
-C_{g',u} c_\b s_W & C_{g,u} c_\b c_W & 0                 & 0 
\end{array}
\right).\label{eq:nkin}
\end{eqnarray}
Mass matrix for $(\tilde B', \tilde W', {\tilde H}_d',{\tilde H}_u')$ is given as,
\begin{eqnarray}
{\cal M}_x &=& ( 1 + \d{\cal N}_{\rm kin.} )^T ( {\cal M}_{\rm MSSM} + \d {\cal M}) ( 1 + \d{\cal N}_{\rm kin.} ).
\end{eqnarray}
Here, ${\cal M}_{\rm MSSM}$ is the neutralino mass matrix in the MSSM, which is given as,
\begin{eqnarray}
{\cal M}_{\rm MSSM} = 
 \left(
  \begin{array}{cccc}
   M_1 & 0 &-m_Z s_W c_\beta & m_Z s_W s_\beta
   \\
   0 & M_2 & m_Z c_W c_\beta &-m_Z c_W s_\beta
   \\
  -m_Z s_W c_\beta & m_Z c_W c_\beta & 0 & -\mu
   \\
   m_Z s_W s_\beta &-m_Z c_W s_\beta & -\mu & 0
  \end{array}
 \right), \label{eq:MneuMSSM}
\end{eqnarray}
and $\d{\cal M}$ is the contributions of the first three terms in Eqs. (\ref{eq:op1}, \ref{eq:op2}),
which is given as,
\begin{eqnarray}
\d {\cal M} \simeq 
 \frac{\e_u \e_d m_Z}{\L} \left(
  \begin{array}{cccc}
   0 & 0 & -2\m{\tilde C}_{g'} s_W c_\b& 2\m{\tilde C}_{g'} s_W s_\b\\
   0 & 0 & 2\m{\tilde C}_{g} c_W c_\b & -2\m{\tilde C}_g c_W s_\b\\
   -2\m{\tilde C}_{g'} s_W c_\b & 2\m{\tilde C}_g c_W c_\b & 0 & - {\bar C} m_Zc_{2\b}/2 \\
   2\m{\tilde C}_{g'} s_W s_\b & -2\m{\tilde C}_g c_W s_\b & -{\bar C}m_Zc_{2\b}/2 & 0
  \end{array}
 \right), \label{eq:deltaM}
\end{eqnarray}
where ${\tilde C}_{g'} = C_{g',d} + C_{g',u}$, ${\tilde C}_{g} = C_{g,d} + C_{g,u}$
and $\bar C = (C_{g',u}-C_{g',d})s_W^2 + (C_{g,u}-C_{g,d})c_W^2$, respectively.
By using Eqs. (\ref{eq:nkin}, \ref{eq:MneuMSSM}, \ref{eq:deltaM}),
we can get the elements of ${\cal M}_x$ as follows:
{\small
\begin{eqnarray}
{\cal M}_x \simeq
 \left(
  \begin{array}{cccc}
   M_1 & 0 &-m_Z s_W c_\beta (1+\d_{13}) & m_Z s_W s_\beta(1+\d_{14})
   \\
   0 & M_2 & m_Z c_W c_\beta (1+\d_{23})&-m_Z c_W s_\beta(1+\d_{24})
   \\
  -m_Z s_W c_\beta (1+\d_{13})& m_Z c_W c_\beta (1+\d_{23})& 0 & -\mu_{\rm eff.}
   \\
   m_Z s_W s_\beta (1+\d_{14})&-m_Z c_W s_\beta (1+\d_{24})& -\mu_{\rm eff.} & 0
  \end{array}
 \right),
\end{eqnarray}
where $\m_{\rm eff.}$ and $\d$'s are given as,
\begin{eqnarray}
\m_{\rm eff.} ~&=&~ -\m -\frac{\e_u \e_d {\bar C}m_Z^2 c_{2\b}}{2\L},\\
\d_{13} ~&=&~ \frac{2C_{g',d}\e_u\e_d }{\L} \left( -M_1 t_\b      + \m \right),\qquad
\d_{14} ~=~ \frac{2C_{g',u}\e_u\e_d }{\L} \left( -M_1 t_\b^{-1} + \m \right),\\
\d_{23} ~&=&~ \frac{2C_{g,d}\e_u\e_d }{\L} \left( -M_2 t_\b      + \m \right),\qquad
\d_{24} ~=~ \frac{2C_{g,u}\e_u\e_d }{\L} \left( -M_2 t_\b^{-1} + \m \right).
\end{eqnarray}

Canonically normalized mass eigenstates $\tilde\chi_i$'s are given as,
\begin{eqnarray}
\left(
\begin{array}{c}
{\tilde\chi}_1^0 \\
{\tilde\chi}_2^0 \\
{\tilde\chi}_3^0 \\
{\tilde\chi}_4^0 \\
\end{array}
\right)
=
{\cal N}
\left(
\begin{array}{c}
{\tilde B}' \\
{\tilde W}' \\
{\tilde H}_d' \\
{\tilde H}_u'  \\
\end{array}
\right)
=
{\cal N}\left( 1+\d {\cal N}_{\rm kin.} \right)^{-1}
\left(
\begin{array}{c}
{\tilde B} \\
{\tilde W}^0 \\
{\tilde H}_d \\
{\tilde H}_u \\
\end{array}
\right).
\end{eqnarray}
We define mixing matrix as ${\cal N}_x \equiv {\cal N}\left( 1+\d {\cal N}_{\rm kin.} \right)^{-1}$.
Mass eigenstate $\tilde\chi_i^0$ can be expressed as,
\begin{eqnarray}
\tilde\chi_i^0 = {\cal N}_{x,1i} \tilde B +{\cal N}_{x,2i} \tilde W +{\cal N}_{x,3i} {\tilde H}_d +{\cal N}_{x,4i} {\tilde H}_u.
\qquad (i=1,2,3,4)
\end{eqnarray}

Hereafter, we assume $m_Z \ll |M_1|-|\m_{\rm eff.}|, |M_2|-|\m_{\rm eff.}|$, for example.
In this case, the lightest neutralino $\tilde\chi_1^0$ and the second lightest neutralino $\tilde\chi_2^0$ is pure higgsino at order $m_Z/M_{1,2}$.
If we take $M_1,~M_2,~\m_{\rm eff.} > 0$,
\begin{eqnarray}
{\cal N}_{x,13} \simeq \frac{1}{\sqrt{2}}, \qquad {\cal N}_{x,14} \simeq \frac{1}{\sqrt{2}}, \qquad
{\cal N}_{x,23} \simeq \frac{1}{\sqrt{2}}, \qquad {\cal N}_{x,24} \simeq -\frac{1}{\sqrt{2}}.
\end{eqnarray}
Bino and wino components in $\tilde\chi_1^0$ and $\tilde\chi_2^0$ are given as,
\begin{eqnarray}
{\cal N}_{x,11} \simeq - \frac{1}{\sqrt{2}}\frac{ {\cal M}_{x,31}+{\cal M}_{x,41} }{M_1-\m_{\rm eff.}} -\d{\cal N}_{ {\rm kin.},31 },\qquad
{\cal N}_{x,12} \simeq - \frac{1}{\sqrt{2}}\frac{ {\cal M}_{x,32}+{\cal M}_{x,42} }{M_2-\m_{\rm eff.}} -\d{\cal N}_{ {\rm kin.},32 },\\
{\cal N}_{x,21} \simeq - \frac{1}{\sqrt{2}}\frac{ {\cal M}_{x,31}-{\cal M}_{x,41} }{M_1+\m_{\rm eff.}} -\d{\cal N}_{ {\rm kin.},41 },\qquad
{\cal N}_{x,22} \simeq - \frac{1}{\sqrt{2}}\frac{ {\cal M}_{x,32}-{\cal M}_{x,42} }{M_2+\m_{\rm eff.}} -\d{\cal N}_{ {\rm kin.},42 }.
\end{eqnarray}
When $|M_1|,|M_2| \gg |\m_{\rm eff}|$, we can get more simpler formulae for ${\cal N}$'s by using the explicit expression of ${\cal M}_x$ and $\d{\cal N}_{\rm kin.}$:
\begin{eqnarray}
{\cal N}_{x,11} &\simeq& \frac{m_Z s_W}{\sqrt{2}}\left[ \left( \frac{c_\b}{M_1} - 2C_{g',d}s_\b\frac{\e_u\e_d}{\L}\right) - \left(\frac{s_\b}{M_1} - 2C_{g',u}c_\b\frac{\e_u\e_d}{\L}\right) \right],\\
{\cal N}_{x,12} &\simeq& \frac{m_Z c_W}{\sqrt{2}}\left[ -\left( \frac{c_\b}{M_2} - 2C_{g,d}s_\b\frac{\e_u\e_d}{\L}\right) + \left(\frac{s_\b}{M_2} - 2C_{g,u}c_\b\frac{\e_u\e_d}{\L}\right) \right],\\
{\cal N}_{x,21} &\simeq& \frac{m_Z s_W}{\sqrt{2}}\left[ \left(\frac{c_\b}{M_1} - 2C_{g',d}s_\b\frac{\e_u\e_d}{\L}\right) + \left(\frac{s_\b}{M_1} - 2C_{g',u}c_\b\frac{\e_u\e_d}{\L}\right) \right],\\
{\cal N}_{x,22} &\simeq& \frac{m_Z c_W}{\sqrt{2}}\left[ -\left(\frac{c_\b}{M_2} - 2C_{g,d}s_\b\frac{\e_u\e_d}{\L}\right) - \left(\frac{s_\b}{M_2} - 2C_{g,u}c_\b \frac{\e_u\e_d}{\L}\right) \right].
\end{eqnarray}

Finally, we mention that, although we only discuss the contribution from the operators in Eqs. (\ref{eq:operators}), there are also contributions from other operators and it have been discussed, for example, in Refs. \cite{Antoniadis:2008es,Berkooz:2008gw,Carena:2009gx,Antoniadis:2009rn,Antoniadis:2010nb,Cassel:2011zd}
\footnote{
For studies of higher dimensional operators in low-energy SUSY models, see also, e.g., Refs. \cite{Cheung:2009qk,Berg:2009mq,Bernal:2009hd,Bernal:2009jc} and 
\cite{Piriz:1997id,Strumia:1999jm,Brignole:2003cm,Casas:2003jx,Pospelov:2005ks,Pospelov:2006jm,Dine:2007xi,Antoniadis:2007xc,Batra:2008rc,Cassel:2009ps,Blum:2009na,Ham:2009gu,Cassel:2010px,Blum:2010by,Bae:2010ai,Franceschini:2010qz,Carena:2010cs,Mukhopadhyay:2011rp,Bernal:2011pj,Altmannshofer:2011rm,Altmannshofer:2011iv,Carena:2011dm,Boudjema:2011aa,Boudjema:2012cq,Boudjema:2012in,Farakos:2012fm,Dudas:2012fa,Berg:2012cg}.}.
%



\end{document}